\documentclass[a4paper,11pt]{article}
\usepackage[utf8]{inputenc}
\usepackage{color}
\usepackage{cite}
\usepackage{hyperref}
\usepackage{amsmath}
\usepackage{graphicx}

\usepackage[top=0.4in, bottom=0.7in, left=0.60in, right=0.5in]{geometry}

\begin{document}

\title{{\Large $J/\Psi$ and $\chi_{cJ}(J=0,1)$  production in electron-positron annihilation at $\sqrt{s}=10.6$ GeV in the framework of Bethe-Salpeter equation}}
\author{Shashank Bhatnagar, Teresa Aruja, Vaishali Guleria}
\maketitle \small{Department of Physics, University Institute of Sciences, Chandigarh University, Mohali-140413, India\\}

\begin{abstract}
\normalsize{In present work we study the production of ground and excited charmonium pairs in $e^- e^+ \rightarrow \Psi(nS)+ \chi_{cJ}(nP)$ for $J=0,1$ and $n=1,2$, through leading order (LO) tree-level diagrams $\sim O(\alpha_{em} \alpha_s)$, which proceed through exchange of a virtual photon and an internal gluon line connecting two quark lines (in the triangle quark loop part of the diagram), in the framework of $4\times 4$ Bethe-Salpeter equation, at center of mass energy, $\sqrt{s}=10.6 GeV.$ We have tried to show that the cross sections for this process calculated with use of the four leading order $\sim O(\alpha_{em} \alpha_s)$ diagrams using BSE approach come close to experimental data, which might be mainly due to the consistent treatment of motion of quarks inside the hadrons in the BSE framework.}
\end{abstract}

\section{Introduction}
In recent years there has been observation of double charm quark pair production by BABAR and Belle Collaborations \cite{belle10,babar09,ulgov05} in $e^- e^+$ annihilation at center of mass energies, $\sqrt{s}=10.6$ GeV. This process is particularly interesting in testing quarkonium production mechanism. This process proceeds through the exchange of a single photon. Charge conjugation symmetry requires that if one of the charmonia is $1^{--}$ (such as $J/|Psi$), then the other should have $C=+$ such as $0^{++} (\chi_{c0})$, $1^{++} (\chi_{c1})$, $2^{++} (\chi_{c2})$, or $0^{-+} (\eta_c)$. In this work we will focus on the two processes: $e^- + e^+ \rightarrow J/\Psi + \chi_{c0}$, and $e^- + e^+ \rightarrow J/\Psi + \chi_{c1}$.

But there is a significant discrepancy between the experimental measurements of total cross sections of $e^- e^+ \rightarrow J/\Psi \eta_c$ and $e^- e^+ \rightarrow J/\Psi \chi_{cJ}$ at energy, $\sqrt{s}=10.6$GeV., and their NRQCD predictions using leading order (LO) diagrams. It is to be mentioned that the LO diagrams alone give cross-sections which are much less than data. Such a large discrepancy between experimental results and theoretical predictions has been a challenge to the understanding of charmonium production through NRQCD.

Many studies have been done to resolve this problem. For instance Ref.\cite{braaten} showed that to obtain cross-sections from NRQCD which are consistent with data, one has to incorporate not only the NLO QCD corrections, (where the value of total NLO contribution to cross section is nearly twice the LO contribution), but also incorporate relativistic corrections.  Incorporation of NLO QCD corrections and relativistic corrections brings their results close to the lower bound set by the Babar and Belle collaborations for the double-charmonia production.

Further, it was recently pointed out that for both the processes, $e^- e^+ \rightarrow J/\Psi \eta_c$ and $e^- e^+ \rightarrow J/\Psi \chi_{cJ}$  in addition to the NLO QCD corrections, the interference between the QCD and QED tree-level diagrams\cite{sun18,sun18a}, namely the $O(\alpha_{em}^2\alpha_s)$-order terms, can also provide signiﬁcant contributions. Their studies have further shown that the colour octet contributions to $e^- e^+\rightarrow J/\Psi \eta_c$, and $e^- e^+ \rightarrow J/\Psi \chi_{cJ}$ are negligibly small, with the dominant contributions \cite{sun18a,sun18} coming from the colour singlet channels.

Since it is seen that the perturbative QCD calculations fail at low energy range typically below the $J/\Psi$ mass scale, thus, at this scale we can employ the Bethe-Salpeter (BS) framework \cite{shi21,guo08,elias11}, which deals with the relativistic motion of quarks inside the hadrons. Thus, in this work, we calculate the cross section for the process taking the leading order (tree-level) diagrams $\sim O(\alpha_{em}\alpha_s)$ in QED and QCD for $J/\Psi$ and $\chi_{cJ}$, $(J=0,1)$ production in $e^- e^+$ annihilation through a virtual photon, and a gluon being exchanged between two quark lines in the quark-triangle diagram shown in Fig.1. In this work we focus only on the colour singlet contributions, which is the dominant channel \cite{sun18}, and derive analytical expressions for their cross sections in the framework of Bethe-Salpeter equation, which is an equation that is firmly rooted in field theory and is supposed to incorporate all the relativistic effects in the bound states of $c\bar{c}$ systems.

Thus, in this work we study the production of ground and excited charmonium meson pairs in $e^- e^+$ annihilation, such as: $J/\Psi(1S)+~ \chi_{c0}(1P);~ \Psi(2S)~+ \chi_{c0}(1P); ~J/\Psi(1S)+ ~\chi_{c0}(2P);~ \Psi(2S)~+ \chi_{c0}(2P)$ along with $e^- + e^+ \rightarrow J/\Psi(1S)~+ \chi_{c1}(1P);~ \Psi(2S)~ +\chi_{c1}(1P); ~J/\Psi(1S)+ ~\chi_{c1}(2P);~ \Psi(2S)~ +\chi_{c1}(2P)$ through leading order (LO) tree-level diagrams $\sim O(\alpha_{em} \alpha_s)$, in the framework of the Bethe-Salpeter equation at center of mass energy, $\sqrt{s}=10.6 GeV.$

We wish to mention that Bethe-Salpeter Equation (BSE) \cite{guo08,hluf19,elias11,shi21} is a conventional approach in dealing with relativistic bound state problems. Due to its firm base in quantum field theory and being a dynamical equation based approach, it provides a realistic description for analyzing hadrons as composite objects and can be applied to study not only the low energy hadronic processes but also the high energy production processes involving quarkonia as well. To simplify the calculations, we make use of the heavy quark limit on the propagators of quark as shown in the next section.

The paper is organised as follows: Section 2 deals with the BS equation for $Q\bar{Q}$ system, Section 3 deals with calculation of cross section for the process, $e^- + e^+ \rightarrow J/\Psi + \chi_{c0}$, while section 4 deals with the corresponding calculation for the process, $e^- + e^+ \rightarrow J/\Psi + \chi_{c1}$. Section 5 deals with the Discussions.

\section{Bethe-Salpeter equation for $Q\bar{Q}$ bound state}
A quark-anti quark bound state system can be described by a Bethe-Salpeter equation (BSE),
\begin{equation}
S_{F}^{-1}(p_{1})\Psi(P,q)S_{F}^{-1}(-p_{2}) =
i\int \frac{d^{4}q'}{(2\pi)^{4}}K(q,q')\Psi(P,q'),
\end{equation}

where $p_1$ and $p_2$ are the momenta of the two particles, with the internal momentum of the hadron being, $q$, and external hadron momentum $P$, and mass, $M$. In Eq.(1), $K(q,q')$ is the interaction kernel, and $S_{F}^{-1}(\pm p_{1,2})=\pm i{\not}p_{1,2}+ m_{1,2}$ are the inverse propagators for the quark and antiquark.

To reduce the above equation to 3D form, we make use of the Covariant Instantaneous Ansatz on the BS kernel, $K(q,q')$, where we can write
$K(q,q')=K(\widehat{q},\widehat{q}')$, where the BS kernel depends entirely on the component of
internal momentum of the hadron, $\widehat{q}_\mu= q_\mu- \frac{q.P}{P^2}P_\mu$, which is a 3D variable, and is orthogonal to the total
hadron momentum, i.e. $\widehat{q}.P=0$, while $\sigma
P_\mu=\frac{q.P}{P^2}P_\mu$ is the component of $q$ that is longitudinal
to $P$. And the 4-dimensional volume element is,
$d^4q=d^3\widehat{q}Md\sigma$.

It is to be observed that the longitudinal component,  $M\sigma'$  of $q'$ does not appear in $K(\hat{q}, \hat{q}')$. We thus carry out integration over the longitudinal component, $Md\sigma'$ of the four dimensional volume element, $d^4q'$ on the right side of Eq.(1). Here, we use of the fact that

\begin{equation}
\psi(\hat{q}')=\frac{i}{2\pi}\int Md\sigma' \Psi(P,q'),
\end{equation}

leading to equation,

\begin{equation}
S_{F}^{-1}(p_{1})\Psi(P,q)S_{F}^{-1}(-p_{2})=\int \frac{d^3\hat{q}'}{(2\pi)^3}K(\hat{q}, \hat{q}')\psi(\hat{q}')=\Gamma(\hat{q}),
\end{equation}
where, $\Gamma(\hat{q})$ is the hadron-quark vertex function, and is directly related to the 4D wave function, $\Psi(P,q)$. Multiplying the previous equation from the left by $S_F(p_1)$, and from the right by $S_F(-p_2)$, we one can express the 4D BS wave function $\Psi(P,q)$ in terms of $\Gamma(\hat{q})$ as,

\begin{equation}
 \Psi(P, q)=S_1(p_1)\Gamma(\hat q)S_2(-p_2),
\end{equation}

where, the 4D hadron-quark vertex, that enters into the definition of the 4D BS wave function in the previous equation is,
\begin{equation}\label{6a}
 \Gamma(\hat q)=\int\frac{d^3\hat q'}{(2\pi)^3}K(\hat q,\hat q')\psi(\hat q').
\end{equation}

Following a sequence of steps we get four Salpeter equations which are the
effective 3D forms of BSE (Salpeter equations)\cite{eshete19}:

\begin{eqnarray}
 &&\nonumber(M-\omega_1-\omega_2)\psi^{++}(\hat{q})=\Lambda_{1}^{+}(\hat{q})\Gamma(\hat{q})\Lambda_{2}^{+}(\hat{q})\\&&
   \nonumber(M+\omega_1+\omega_2)\psi^{--}(\hat{q})=-\Lambda_{1}^{-}(\hat{q})\Gamma(\hat{q})\Lambda_{2}^{-}(\hat{q})\\&&
\nonumber \psi^{+-}(\hat{q})=0.\\&&
 \psi^{-+}(\hat{q})=0\label{fw5}
\end{eqnarray}

These four Salpeter equations were used to derive the mass spectral equations of $0^{++}, 0^{-+}, 1^{--}, 1^{++}$ and $1^{+ -}$ mesons\cite{eshete19,vaishali21a}. The solutions of these mass spectral equations not only lead to their mass spectrum but also the analytical structures of their radial wave functions, that were used for calculations of various transitions such as decay constants of these quarkonia, as well as $M1$ and $E1$ transitions such as, $V\rightarrow P\gamma, P\rightarrow V\gamma$, and $V\rightarrow S\gamma, S\rightarrow V\gamma, A^-\rightarrow P\gamma, P\rightarrow A^-\gamma$ respectively \cite{bhatnagar20,vaishali21}. The analytic structures of these radial wave functions of $S,V$ and $A^+$ mesons that are used for study of charmonium production processes: $e^- +e^+\rightarrow J/\Psi +\chi_{cJ} (J=0,1)$ in the present work, are given in Eqs.(12), and (23).

Thus, it is to be mentioned that in our framework, the component, $\hat{q}_{\mu}$, is always orthogonal to $P_{\mu}$ and satisfies the relation, $\hat{q}.P=0$, irrespective of whether $q.P=0$ (i.e. $\sigma=0$), or $q.P\neq 0$ (i.e. $\sigma \neq 0$). Further, due to the Lorentz-invariant nature of $\hat{q}^2$, the applicability of this framework of Covariant Instantaneous Ansatz is valid all the way from low energy spectra to high energy transition amplitudes\cite{eshete19,vaishali21}.

The interaction kernel in BSE is taken as a one-gluon-exchange type as regards the colour and spin dependence, and thus has a general structure \cite{eshete19}, $K(\hat{q},\hat{q}')=(\frac{1}{2}\lambda_1.\frac{1}{2}\lambda_2)\gamma_{\mu}\times \gamma_{\mu} V(\hat{q}, \hat{q}')$, where the scalar part of the kernel is written as a sum of the one-gluon-exchange part and the confinement part as, $V(\hat{q},\hat{q}')= \frac{4\pi\alpha_s}{(\hat q-\hat
q')^2}
 +\frac{3}{4}\omega^2_{q\bar q}\int d^3r\bigg(\kappa r^2-\frac{C_0}{\omega_0^2}\bigg)e^{i(\hat q-\hat q').\vec r}$, and
$ \kappa=(1+4\hat m_1\hat m_2A_0M^2r^2)^{-\frac{1}{2}}$ (see\cite{eshete19} for details).

Regarding the parameters of the model, $\omega_0=0.22$ GeV. is the spring constant, $C_0=0.69$ is a dimensionless constant, while $\frac{C_0}{\omega_0^2}$ plays the role of  ground state energy, $\Lambda=0.0250$ GeV. is the QCD length scale,  with input quark masses, $m_u= 0.300 GeV, m_s= 0.430 GeV, m_c= 1.490 GeV$, and $m_b= 4.690$ GeV. Our previous studies on mass spectral calculations of heavy-light quarkonia\cite{eshete19} were used to fit the input parameters of our model. The cross sections are calculated with the inverse range parameter $\beta=0.0900$GeV.

\section{Cross section for $e^- + e^+ \rightarrow J/\Psi + \chi_{c0}$}

The four colour singlet leading-order (LO) Feynman diagrams $ O(\alpha_{em} \alpha_s)$ for the process, $e^- +e^+ \rightarrow J/\Psi+\chi_{c0}$, two of which are given in Figure 1. These diagrams along with two other exchange diagrams form a gauge invariant subset. In all these diagrams, the coupling of virtual photon to the $c\bar{c}$ mesons is described by the quark loop diagrams, where the photon-quark-anti-quark vertex is given as $ie_Q\gamma_{\mu}$, where, $e_Q=\frac{2}{3}e$ is the charge of the $c$ quark, and there is a single gluon exchange between two quark lines. Since the amplitude for the process from all diagrams will be equal, we can express the total amplitude for the process as four times the amplitude from the diagram in Fig.2.
\begin{figure}[h!]
 \centering
 \includegraphics[width=15cm,height=6cm]{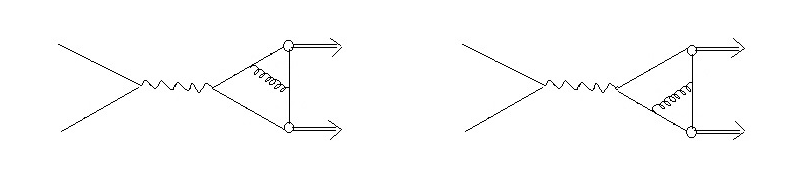}
 \caption{Two of the lowest order Feynman diagrams for the production of a pair of charmonia in electron-positron annihilation. Other two diagrams can be obtained by permutations}
\end{figure}

\begin{figure}[h!]
 \centering
 \includegraphics[width=10cm,height=6cm]{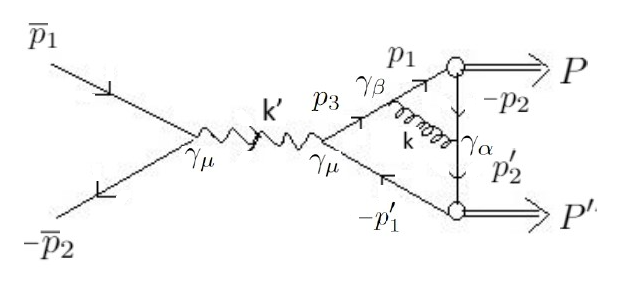}
 \caption{Momentum labeling of the ﬁrst of the two Feynman diagrams shown in Fig. 1}
 \end{figure}

The invariant amplitude $M^{1}_{fi}$ for double charmonium production, corresponding to Fig.2, is given by the
one-loop momentum integral as:
\begin{equation}
M^I_{fi}=e e_Q g^2F_{12}[\bar{v}^{(s2)}({\bar{p}}_2)\gamma_{\mu}u^{(s1)}(\bar{p}_1)]\frac{-1}{(\bar{p}_1+\bar{p}_2)^2}\int \frac{d^4q}{(2\pi)^4}\int \frac{d^4q'}{(2\pi)^4}Tr[\bar{\Psi}_V(P,q)\gamma_{\beta}S_F(p_3)\gamma_{\mu}\bar{\Psi}_S(P', q')\gamma_{\alpha}]\frac{\delta_{\alpha\beta}}{k^2},
\end{equation}

where, $P, q$ and $\epsilon^{\lambda}$ are the external momentum, internal momentum and polarization vector of $J/\Psi$, while, $P', q'$ are corresponding variables of $\chi_{c0}$.  $F_{12}=\frac{1}{2}\vec{\lambda_1}.\frac{1}{2}\vec{\lambda_1}=\frac{4}{3}$ is the colour factor, and the wave functions of both incoming and outgoing particles are normalized to one particle per unit volume. The above expression can thus be factorized as the scalar product of the leptonic current and the hadronic current as, $M^I_{fi}=J_{\mu}^{em} J_{\mu}^{hadronic}$. The total amplitude for the process, $M_{fi}$ will be four times the amplitude, $M^I_{fi}$ for diagram 1. The 4D adjoint BS wave functions of V and S mesons can be written as,
\begin{eqnarray}
&&\nonumber \bar{\Psi}_V(P, q)=S_F(-p_2)\Gamma_V(\hat{q})S_F(p_1)\\&&
\bar{\Psi}_S(P', q')=S_F(-p'_1)\Gamma_S(\hat{q}')S_F(p'_2)
\end{eqnarray}

where $-p_2 = -p'_2$. Here, we consider $p_{1,2}$ as the momenta of quark/anti-quark of the vector meson, $J/\Psi$ of total momentum, $P$, while $p'_{1,2}$ as the momenta of the quark/anti-quark of the scalar ($\chi_{c0}$)/ axial vector ($\chi_{c1}$) meson of momentum $P'$. These quark momenta can be expressed in terms of the total and the internal momenta of respective mesons as,
\begin{equation}
p_1=\frac{1}{2}P + q,~~~~ p_2=\frac{1}{2}P -q,~~~~p'_1 =\frac{1}{2}P' -q',~~~~~~~ p'_2=\frac{1}{2}P'+q'.
\end{equation}

To simplify the calculation, we assume that the propagators of the quark and the gluon are independent of relative momenta $q$, and $q'$ (see \cite{guo08,elias11,hluf19}). This simplification is appropriate since the masses of heavy quarks are large compared with the relative momenta, which are of order $\alpha_s m_c$. Thus the momenta $p_3$ and $k$ of the propagators are large in comparison with the relative momenta, $q$ and $q'$ of the mesons. One may expect that, in the heavy quark limit, the calculation without taking into account the relative momenta should be a good approximation.

Thus under heavy quark approximation, we can take, $p_1=p_2\approx \frac{1}{2}P$, and $p_1' = p_2'\approx \frac{1}{2}P'$. Thus we can write the momentum of the internal gluon propagator as, $k=\bar{p_1}+\bar{p_2}\approx \frac{1}{2}(P+P')$. Thus $k^2\approx -\frac{s}{4}$. Similarly, we can express the momentum $p_3$ of one of the internal quark lines as, $p_3=p_1+k=\frac{1}{2}P'+P+q \approx \frac{1}{2}P'+P$. Similarly, $p_3^2$ that enters into the denominator of this propagator can be expressed as, $p_3^2\approx -\frac{s}{2}-M^2-\frac{M'^2}{4}$.

Now,we make use of the fact that we decompose $q=(\hat{q},M\sigma)$, and $q'=(\hat{q}', M'\sigma')$, while $e_Q=\frac{2}{3}e$ is the charge of the charmed quark. Further we can express the four-dimensional volume elements,
$d^4q= d^3\hat{q} Md\sigma$, and $d^4 q'= d^3\hat{q}' M'd\sigma'$, and making use of the fact that,

\begin{eqnarray}
&&\nonumber \int \frac{Md\sigma}{2\pi}\Psi_V(P,q)= \Psi_V(\hat{q})\\&&
\int \frac{M'd\sigma'}{2\pi}\Psi_S(P',q')= \Psi_S(\hat{q}').
\end{eqnarray}

Now, to calculate $M_{fi}$, we use the 3D decomposition of the general form of 4D BS wave functions \cite{smith69} for vector and scalar mesons.  Further, making use of the most dominant Dirac structures for both vector and scalar mesons, that contribute maximum to calculation of any meson observable\cite{bhatnagar06,bhatnagar11,bhatnagar14}, we take the structures of 3D wave functions as \cite{bhatnagar20}:

\begin{equation}\label{wf44}
\begin{split}
  \psi_S(\hat q')=N_S(M')\phi_S(\hat q')\\
 \psi_V(\hat q)=N_V(iM{\not}\epsilon)\phi_V(\hat q),
 \end{split}
\end{equation}

where the scalar functions, $\phi_V(\hat q)$, and $\phi_S(\hat q')$ are obtained as analytic solutions of mass spectral equations for vector and scalar mesons, that are in turn obtained through solutions of their corresponding mass spectral equations\cite{eshete19}. These radial wave functions for the ground and excited states of vector and scalar mesons are\cite{eshete19}:
\begin{eqnarray}
&&\nonumber \phi_S(1P,\hat q)=\sqrt{\frac{2}{3}}\frac{1}{\pi^{3/4}}\frac{1}{\beta_S^{5/2}} \hat q e^{-\frac{\hat q^2}{2\beta_S^2}},\\&&
\nonumber \phi_S(2P,\hat q)=\sqrt{\frac{5}{3}}\frac{1}{\pi^{3/4}}\frac{1}{\beta_S^{5/2}}\hat q(1-\frac{2\hat q^2}{5\beta_S^2})e^{-\frac{\hat q^2}{2\beta_S^2}},\\&&
\nonumber \phi_V(1S,\hat q)=\frac{1}{\pi^{3/4}\beta_V^{3/2}} e^{-\frac{\hat q^2}{2\beta_V^2}},\\&&
 \phi_V(2S,\hat q)=\sqrt{\frac{3}{2}}\frac{1}{\pi^{3/4}\beta_V^{3/2}}(1-\frac{2\hat q^2}{3\beta_V^2}) e^{-\frac{\hat q^2}{2\beta_V^2}},
\end{eqnarray}

where $\beta_v$ and $\beta_s$ are their inverse range parameters. To evaluate the trace over the gamma matrices, we need the expressions for the adjoint 3D wave functions, $\bar{\Psi}_V(\hat{q})$ and $\bar{\Psi}_V(\hat{q})$, where we express $\bar{\Psi}=\gamma_4 \Psi^\dag \gamma_4$.
We then write the invariant amplitude $M_{fi}$ in the center of mass frame (where, $\vec{\bar{p_1}}=- \vec{\bar{p_2}}$, and   $E_1=E_2 (=E))$, and with
$\sqrt{s}=2E$, after evaluation of trace as,

\begin{equation}
M_{fi}=i\frac{128}{3}\frac{ee_Q g^2}{s} [\bar{v}^{(s2)}({\bar{p}}_2){\not} \epsilon u^{(s1)}(\bar{p}_1)]\frac{m_c N_V N_S M M'}{k^2(-\frac{s}{2}-M^2-\frac{M'^2}{4}+m_c^2)}\int \frac{d^3\hat{q}}{(2\pi)^3}\phi_V(\hat{q})\int \frac{d^3\hat{q}'}{(2\pi)^3}\phi_S(\hat{q}')
\end{equation}

Now, to evaluate spin averaged invariant amplitude modulus squared, $|\bar{M}_{fi}|^2$, we need to average over the initial spin states of electron and positron, and sum over the polarization states of vector meson, that is, $|\bar{M}_{fi}|^2=\frac{1}{4}\sum_{s_1,s_2,\lambda} M^\dag_{fi} M_{fi}$, which after evaluation of trace over gamma matrices and making use of the normalizations,
$\Sigma_{\lambda} \epsilon_{\mu}^{\lambda}\epsilon_{\nu}^{\lambda}=\frac{1}{3}(\delta_{\mu\nu}+\frac{P_{\mu}P_{\nu}}{M^2})$ for polarization vectors of the vector meson can be expressed as,

\begin{eqnarray}
&&\nonumber |\bar{M}_{fi}|^2=\frac{2^{22}}{3^4}\frac{\pi^4 \alpha_{em}^2\alpha_s^2}{s^4} \bigg[-\frac{1}{3}\bar{p}_1.\bar{p}_2+\frac{2}{3}\frac{(\bar{p}_1.P)(\bar{p}_2.P)}{M^2}+m_e^2\bigg]\frac{m_c^2 N_V^2N_S^2 M^2 M'^2}{(-\frac{s}{2}-M^2-\frac{M'^2}{4}+m_c^2)^2}\xi_V^2\xi_S^2\\&&
\nonumber \xi_V=\int \frac{d^3\hat{q}}{(2\pi)^3}\phi_V(\hat{q})\\&&
\xi_S=\int \frac{d^3\hat{q'}}{(2\pi)^3}\phi_S(\hat{q}'),
\end{eqnarray}

which has the general form,
\begin{equation}
|\bar{M}_{fi}|^2=\frac{2^{22}}{3^4}\frac{\pi^4 \alpha_{em}^2\alpha_s^2}{s^4}\frac{m_c^2 N_V^2N_S^2 M^2 M'^2}{(-\frac{s}{2}-M^2-\frac{M'^2}{4}+m_c^2)^2}\xi_V^2\xi_S^2 L_{\mu\mu'}H_{mu\mu'},
\end{equation}

where $L_{\mu\mu'}$ is the leptonic tensor, while $H_{\mu\mu'}$ is the hadronic tensor. Here electromagnetic coupling constant, $\alpha_{em}=\frac{e^2}{4\pi}$, and the strong coupling constant, $\alpha_s=\frac{g^2}{4\pi}=\frac{1}{log(m_c/\Lambda_{QCD})}$, $m_e$ is electron mass, while $m_c$ is the c-quark mass, and we have made use of $k^2=-\frac{s}{4}$. Now, $M_{fi}$ in previous equation involves the dot products of momenta, which can be expressed in the center of mass frame as, $\bar{p}_1.\bar{p}_2=\frac{1}{2}(2m_e^2-s),~~~~ \bar{p}_1.P\approx -E^2(1-cos\theta), ~~~~\bar{p}_2.P\approx -E^2(1+cos\theta)$, with $\sqrt{s}=2E=E_{cm}$, being the center of mass energy. Here, $(\bar{p}_1+\bar{p}_2)=\sqrt{s}$. It can be checked that the asymptotic behaviour of amplitude, $M_{fi}\sim 1/s^3$.

Further, $N_S$ and $N_V$ are the 4D BS normalizers of vector and scalar mesons, that are obtained through the current conservation condition,

\begin{equation}\label{46}
2iP_\mu=\int \frac{d^{4}q}{(2\pi)^{4}}
\mbox{Tr}\left\{\overline{\Psi}(P,q)\left[\frac{\partial}{\partial
P_\mu}S_{F}^{-1}(p_1)\right]\Psi(P,q)S_{F}^{-1}(-p_2)\right\} +
(1\rightleftharpoons2),
\end{equation}

and are expressed as,

\begin{eqnarray}
&&\nonumber N_V^{-2}=4\hat m_1 \widehat m_2 M^2\frac{1}{m_c}\int \frac{d^3\widehat{q}}{(2\pi)^3}\phi_V^2(\widehat{q}),\\&&
 N_{S}^{-2}=4\hat m_1 \widehat m_2 M'^2\frac{1}{m_1}\int \frac{d^3\widehat{q}'}{(2\pi)^3}\phi_{S}^2(\widehat{q}'),
\end{eqnarray}

respectively.

We can express $d\sigma$ in terms of the transition rate per unit volume, $W_{fi}$ as,
\begin{equation}
d\sigma=\frac{W_{fi}(2\pi)^4\delta^{(4)}(p_1+p_2-P'-P'')}{F}[\frac{d^3\vec{P}'}{2E'(2\pi)^3}\frac{d^3\vec{P}''}{2E''(2\pi)^3}],
\end{equation}

where  the factor in the square brackets on the extreme right is the total phase space volume for the two outgoing particles  in the center of mass frame considered. And $F$ is the incident flux, which is expressed as, $F=4\sqrt{(\bar{p_1}.\bar{p_2})^2-m_1^2 m_2^2}$ (which includes the kinematical factors of initial particles), and can in turn be expressed in the Center of mass frame as, $F=4|\vec{p}|\sqrt{s}$, where $|\vec{p}|$ is the magnitude of momentum of either of the incident particles in the center of mass frame, and $\sqrt{s}=2E$ is the center of mass energy of incident particles. Following a series of steps, we can express the total cross section as,

\begin{equation}
\sigma=\frac{1}{32\pi^2 s^{3/2}}|\vec{P'}|\int d\Omega' |\bar{M}_{fi}|^2,
\end{equation}

with $|\vec{P'}|=\sqrt{\frac{1}{s}[s-(M+M')^2][s-(M-M')^2]}$ being the momenta of either of the outgoing particles in the center of mass frame, and $|\bar{M}_{fi}|^2$, being expressed as in Eq.(7). As regards the behaviour of various quantities, it can be checked that $\beta_S\sim \beta_V \sim M$, The BS normalizers behave as, $N_S \sim N_V \sim M^{-1/2}$, while the 3D integrals, $\xi_S \sim \xi_V \sim M^{3/2}$. Thus the cross sectional formula behaves as, $\sigma \sim M^{-2}$.

Thus we have expressed the amplitude and the cross section for the process, $e^+ e^-\rightarrow J/\Psi \chi_{c0}$ in terms of the radial wave functions of both the produced vector and scalar quarkonia, which were analytically obtained from solutions\cite{eshete19} of mass spectral equations of vector and scalar mesons. The cross sections for $e^-~~ e^+\rightarrow \Psi(nS)~~ \chi_{c0}(nP)$ calculated in this work are given in Table 1 along with experimental data and results of other models. It is seen that the contribution of wave functions at the origin $\xi_V$ and $\xi_S$ have a major role to play in the calculation of the cross section. Their numerical values are: 0.0308 GeV. and 0.0336 GeV. for $J/Psi$ and $\chi_{c0}(1P)$ respectively.

\begin{table}[hhhhh]
  \begin{center}
  \begin{tabular}{p{4.5cm} p{2.8cm} p{2.8cm} p{2.5cm} p{2.2cm} p{1.8cm} }
  \hline
 Process & BSE-CIA& Expt. \cite{ulgov05}& \cite{sun18}&\cite{braaten}&\cite{chao}\\
  \hline
  $e^- e^+\rightarrow J/\Psi(1S) \chi_{c0}(1P)$  &  6.771& 17$\pm$ 8$\pm$7&6.62&2.40$\pm$1.02&5 \\
  $e^- e^+\rightarrow \Psi(2S) \chi_{c0}(1P)$     &  3.639 & &&1.00$\pm$0.42&\\
  $e^- e^+\rightarrow \Psi(1S) \chi_{c0}(2P)$ &    2.950& & & &\\
  $e^- e^+\rightarrow \Psi(2S) \chi_{c0}(2P)$ & 3.176 &  & & &\\
   \hline
  \end{tabular}
\caption{Cross sections for process, $e^-~~ e^+\rightarrow \Psi(nS)~~ \chi_{c0}(nP)$ calculated in BSE-CIA along with experimental data and results of other models in fb}
\end{center}
\end{table}

\section{Cross section for $e^- + e^+ \rightarrow J/\Psi + \chi_{c1}$}
In lowest order the process proceeds through the same Feynman diagrams given in Fig.1, except that now we have $J/\Psi$ and $\chi_{c1}$ in the final state, with external momentum, $P'$, internal momentum, $q'$, and polarization vector, $\epsilon^{\lambda'}$, while the corresponding variables for $J/\Psi$ remain the same as, $P, q$ and $\epsilon^{\lambda}$ as in Diagram 1.

Amplitude for this process can thus be written as,

\begin{equation}
M^I_{fi}=e e_Q g^2F_{12}[\bar{v}^{(s2)}({\bar{p}}_2)\gamma_{\mu}u^{(s1)}(\bar{p}_1)]\frac{-1}{(\bar{p}_1+\bar{p}_2)^2}\int \frac{d^4q}{(2\pi)^4}\int \frac{d^4q'}{(2\pi)^4}Tr[\bar{\Psi}_V(P,q)\gamma_{\beta}S_F(p_3)\gamma_{\mu}\bar{\Psi}_A(P', q')\gamma_{\alpha}]\frac{\delta_{\alpha\beta}}{k^2},
\end{equation}

where, the 4D BS adjoint BS wave functions are:
The 4D adjoint BS wave functions of V and S mesons can be written as,
\begin{eqnarray}
&&\nonumber \bar{\Psi}_V(P, q)=S_F(-p_2)\Gamma_V(\hat{q})S_F(p_1)\\&&
\bar{\Psi}_A(P', q')=S_F(p'_1)\Gamma_A(\hat{q}')S_F(-p'_2)
\end{eqnarray}

We make use of the definitions of quark momenta in Eq.(2),  and make use of the definitions of the quark momenta mentioned in paragraph after Eq.(9). Further, the 4D volume elements, $d^4q= d^3\hat{q} Md\sigma$, and $d^4 q'= d^3\hat{q}' M'd\sigma'$, and we make use of the fact that,

\begin{eqnarray}
&&\nonumber \int \frac{Md\sigma}{2\pi}\Psi_V(P,q)= \Psi_V(\hat{q})\\&&
\int \frac{M'd\sigma'}{2\pi}\Psi_A(P',q')= \Psi_A(\hat{q}').
\end{eqnarray}

Now, to calculate $M_{fi}$, we use the 3D decomposition of the general form of 4D BS wave functions \cite{smith69} for vector and scalar mesons, making use of the most dominant Dirac structures for both vector and scalar mesons, that contribute maximum to calculation of any meson observable, as in \cite{bhatnagar20,vaishali21}, and take,

\begin{eqnarray}
&&\nonumber \psi_V(\hat q)=N_V iM{\not}\epsilon^{\lambda}\phi_V(\hat q)\\&&
\psi_A(\hat q')=N_A i M'\gamma_5 {\not}\epsilon^{\lambda'} \phi_A(\hat q').
\end{eqnarray}

where the ground and excited radial wave functions of axial mesons that are obtained through the solutions of their mass spectral equations are:
\begin{eqnarray}
&&\nonumber \phi_A(1P,\hat q)=\sqrt{\frac{2}{3}}\frac{1}{\pi^{3/4}}\frac{1}{\beta_A^{5/2}} \hat q e^{-\frac{\hat q^2}{2\beta_A^2}},\\&&
\phi_A(2P,\hat q)=\sqrt{\frac{5}{3}}\frac{1}{\pi^{3/4}}\frac{1}{\beta_A^{5/2}}\hat q(1-\frac{2\hat q^2}{5\beta_A^2})e^{-\frac{\hat q^2}{2\beta_A^2}}
\end{eqnarray}

Performing integrals over the longitudinal components, $Md\sigma$ and $M'd\sigma'$ in Eq.(20) through use of Eq.(22), putting the expressions for 3D BS wave functions $\Psi_V(\hat{q})$ and $\Psi_A(\Hat{q}')$,  evaluating trace over the gamma matrices, we can express the amplitude, $M_{fi}$ as,

\begin{eqnarray}
&&\nonumber M_{fi}=\frac{2^9 e e_Q g^2}{3 s^2 (\frac{s}{2}+M^2+M'^2/2-m_c^2)}N_A N_V \xi_V \xi_A[\bar{v}^{(s_2)}(p_2)\gamma_{\mu}u^{(s_1)}(p_1)]\epsilon_{\mu \nu \alpha \beta} \epsilon^{\lambda}_{\nu}\epsilon^{\lambda'}_{\alpha}(P+\frac{1}{2}P')_{\beta};\\&&
\nonumber \xi_V=\int \frac{d^3\hat{q}}{(2\pi)^3}\phi_V(\hat{q})\\&&
\xi_A=\int \frac{d^3\hat{q'}}{(2\pi)^3}\phi_A(\hat{q}'),
\end{eqnarray}

Now, to evaluate spin averaged invariant amplitude modulus squared, $|\bar{M}_{fi}|^2$, we need to average over the initial spin states of electron and positron, and sum over the polarization states of vector meson and axial meson, that is, $|\bar{M}_{fi}|^2=\frac{1}{4}\sum_{s_1,s_2,\lambda,\lambda'} M^\dag_{fi} M_{fi}$, which after evaluation of trace over gamma matrices and making use of the normalizations,
$\Sigma_{\lambda} \epsilon_{\mu}^{\lambda}\epsilon_{\nu}^{\lambda}=\frac{1}{3}(\delta_{\mu\nu}+\frac{P_{\mu}P_{\nu}}{M^2})$ for polarization vectors of the vector meson, and $\Sigma_{\lambda'} \epsilon_{\mu}^{\lambda'}\epsilon_{\nu}^{\lambda'}=\frac{1}{3}(\delta_{\mu\nu}+\frac{P'_{\mu}P'_{\nu}}{M'^2})$ for polarization vectors of the axial meson, can be expressed as,

\begin{eqnarray}
&&\nonumber |\bar{M}_{fi}|^2=-\frac{2^{26}\alpha_{em}^2\alpha_s^2 \pi^4} {81s^4(\frac{s}{2}+M^2+M'^2/2-m_c^2)^2}\epsilon_{\mu \nu\alpha\beta}~~\epsilon_{\mu' \nu'\alpha'\beta'}\bigg(p_{1\mu'}p_{2\mu}+p_{1\mu}p_{2\mu'}+(-p_1.p_2+m_e^2)\delta_{\mu \mu'}\bigg)\\&&
\bigg(\delta_{\nu\nu'}\delta_{\alpha\alpha'}+\frac{P_{\nu}P_{\nu'}}{M^2}\delta_{\alpha\alpha'}+\frac{P_{\alpha}P_{\alpha'}}{M'^2}\delta_{\nu\nu'}+\frac{P_{\nu} P_{\nu'} P'_{\alpha} P_{\alpha'}}{M^2 M'^2}\bigg)(P+\frac{1}{2}P')_\beta (P+\frac{1}{2}P')_\beta' N_A^2 N_V^2 M^2 M'^2 \xi_v^2 \xi_A^2.
\end{eqnarray}

Thus, we express $|\bar{M}_{fi}|^2$ in terms of dot products of momenta in the center of mass frame as, $\bar{p}_1.\bar{p}_2=\frac{1}{2}(2m_e^2-s),~~~~ \bar{p}_1.P\approx -E^2(1-cos\theta), ~~~~\bar{p}_2.P\approx -E^2(1+cos\theta), \bar{p}_1.P'\approx -E^2(1+cos\theta)= \bar{p}_2.P $ as,

\begin{eqnarray}
&&\nonumber |\bar{M}_{fi}|^2=-\frac{2^{28}\alpha_{em}^2\alpha_s^2 \pi^4} {9^3 s^4(\frac{s}{2}+M^2+M'^2/2-m_c^2)^2} N_A^2 N_V^2 M^2 M'^2 \xi_v^2 \xi_A^2 |\overline{TR}|^2;\\&&
\nonumber |\overline{TR}|^2=\frac{1}{128M^2 M'^2} \bigg[4 M^4 \bigg(64 m_e^2 M'^2+ s (-96 M'^2 + 19 s - 7 s \cos^2\theta)\bigg) +\\&&
\nonumber s\bigg(-8M'^6 + 3 s^3 (-5 + \cos^2\theta) + 2 M'^2 s (8 m_e^2 - 5 s + 7 s \cos^2\theta) + M'^4 s (15 + \cos \theta (3 + 4\cos \theta))\bigg) +\\&&
2 M^2 \bigg(32 m_e^2 (M'^4 + 4 M'^2 s + s^2) + s [56 M'^4 + 4 s^2 (3 + \cos^2\theta) + M'^2 s (63 + \cos \theta (-8 + 3\cos \theta))]\bigg)\bigg]
\end{eqnarray}

Thus, the total cross section can then be expressed as,

\begin{equation}
\sigma=\frac{1}{16\pi s^{3/2}}|\vec{P'}|\int d \cos \theta |\bar{M}_{fi}|^2.
\end{equation}

Our results of cross sections for $e^-~~ e^+\rightarrow \Psi(nS)~\chi_{c1}(nP)$  are listed in table 2 along with experimental data and results of other models.

\begin{table}[hhhhh]
  \begin{center}
\begin{tabular}{p{4.5cm} p{2.8cm} p{2.8cm} p{2.5cm} p{2.2cm} p{1.8cm} }
  \hline
 Process & BSE-CIA& Expt. \cite{ulgov05}& \cite{sun18}&\cite{braaten}&\cite{chao}\\
  \hline
  $e^- e^+\rightarrow J/\Psi(1S) \chi_{c1}(1P)$  &  9.446 & $<$ 18 &0.53&0.38$\pm$0.12&18 \\
  $e^- e^+\rightarrow \Psi(2S) \chi_{c1}(1P)$     &  7.535 & && &\\
  $e^- e^+\rightarrow \Psi(1S) \chi_{c1}(2P)$ &    6.577& & & &\\
   $e^- e^+\rightarrow \Psi(2S) \chi_{c1}(2P)$ & 5.199 &  & & &\\
   \hline
  \end{tabular}
\caption{Cross sections for process, $e^-~~ e^+\rightarrow \Psi(nS)~~ \chi_{c1}(nP)$ calculated in BSE-CIA along with experimental data and results of other models in fb}
\end{center}
\end{table}

\section{Discussions}
In this work we study the production of ground and excited charmonium pairs in $e^- e^+ \rightarrow \Psi(nS)+ \chi_{cJ}(nP)$ for $J=0,1$ and $n=1,2)$, through leading order (LO) tree-level diagrams $\sim O(\alpha_{em} \alpha_s)$, which proceed through exchange of a virtual photon and an internal gluon line connecting two quark lines in the triangle quark loop part of the diagram, in the framework of $4\times 4$ Bethe-Salpeter equation under Covariant Instantaneous Ansatz, at center of mass energy, $\sqrt{s}=10.6 GeV.$ One of the biggest challenges in calculations based on NRQCD was that the theoretical predictions of cross sections for the processes, $e^- e^+ \rightarrow J/\Psi+ \chi_{cJ}(nP)$, and $e^- e^+ \rightarrow J/\Psi+ \eta_c$ using leading order QCD diagrams is an order of magnitude lower than the experimental measurements \cite{belle10,babar09,ulgov05} of the same. And it was only after use of NLO QCD diagrams and incorporation of relativistic corrections that the results of cross sections came near the lower bound set by BABAR and Belle\cite{belle10,babar09,ulgov05}.

In these calculations, we try to show that the cross sections for double charmonium production calculated with use of the four leading order $\sim O(\alpha_{em} \alpha_s)$ diagrams using BSE approach come close to experimental data, which might be mainly due to the consistent treatment of motion of quarks inside the hadrons in the BSE framework, which is firmly rooted in field theory, though to simplify evaluation of integrals in amplitude, $M_{fi}$ we have made use of the heavy-quark approximation on the quark propagator, $S_F(p_3)$ and the gluon propagator, where the propagators of the quark and the gluon are independent of internal momenta $q$, and $q'$ \cite{guo08,elias11,hluf19}. This simplification seems appropriate since the masses of heavy quarks are large compared with the relative momenta, which are of order $\alpha_s m_c$. Thus the momenta $p_3$ and $k$ of the propagators can be considered to be large in comparison with the internal momenta of the two mesons.

We do expect that the above heavy-quark approximation on the quark and gluon propagators will not lead to a value of cross section that is drastically different from the calculation where we employ the dependence of quark and gluon momenta on the internal momenta of the two hadrons, which we intend to do next on lines of our previous works on $M1$ and $E1$ decays \cite{bhatnagar20,vaishali21} of quarkonia.

However what is interesting in the present calculation is the simplicity by which one can deduce the main features of these cross section calculations for these interesting processes using the leading order diagrams, $\sim O(\alpha_{em} \alpha_s)$ in QCD in the framework of BSE, namely the consistent treatment of internal motion of quarks in hadrons in BSE framework, which we feel is the single most important reason for obtaining theoretical values of cross sections of the same order of magnitude as the lower bound set by BABAR and Belle\cite{belle10,babar09,ulgov05}. We wish to point out this is further validated by a recent calculation \cite{luchinsky} on $e^- e^+ \rightarrow J/\Psi \eta_c$ using Light cone, where the authors had shown that by taking intrinsic motion of quarks inside the hadrons, one can significantly increase the value of cross section.

This calculation involving production of double charmonia in electron-positron annihilation can be easily extended to studies on other processes (involving the exchange of a single virtual photon) observed at B-factories such as,$e^+ e^- \rightarrow \gamma \chi_{cJ}$, and also to processes proceeding through exchange of two virtual photons such as $e^+ e^- \rightarrow J/\Psi J/\Psi$.

Among other similar works, in \cite{shi21}, the authors study in the framework of Bethe-Salpeter equation, the production ratio of neutral to charged kaon pair in \(e^+ e^-\) annihilation below the \(J \slash \psi\) mass by employing the two leading Dirac structures in hadronic wave functions as per the power counting rule \cite{bhatnagar06,bhatnagar11} we had proposed some time ago. We wish to extend this study to calculation of cross section for each of these processes with the involvement of all the possible Dirac structures.

\end{document}